\newif\ifconf\conffalse

\ifconf
\documentclass[twoside,leqno,twocolumn]{article}
\usepackage{ltexpprt}

\else
\documentclass[11pt,letterpaper]{article}
\usepackage{amsthm}
\usepackage[margin=1in,dvips]{geometry}
\fi

\usepackage{color}
\usepackage{import}
\usepackage{verbatim}
\usepackage{tikz}
\usetikzlibrary{quotes,angles}
\usepackage{graphicx}
\usepackage{amssymb}
\usepackage{amsmath}
\usepackage{float}
\usepackage{enumerate}
\usepackage[boxed,section]{algorithm}
\usepackage{algpseudocode}
\usepackage{multirow}
\usepackage[hidelinks]{hyperref}

\iffalse
\newtheorem{definition}{Definition}[section]
\newtheorem{remark}{Remark}[section]
\else
\newtheorem{theorem}{Theorem}[section]
\newtheorem{lemma}[theorem]{Lemma}

\newtheorem{fact}[theorem]{Fact}
\fi

\newcommand{\abs}[1]{\left|#1\right|}

\newcommand{\floor}[1]{\left\lfloor#1\right\rfloor}
\newcommand{\norm}[2]{\left \lVert#2\right \rVert_{#1}}

{\makeatletter
 \gdef\xxxmark{%
   \expandafter\ifx\csname @mpargs\endcsname\relax 
     \expandafter\ifx\csname @captype\endcsname\relax 
       \marginpar{xxx}
     \else
       xxx 
     \fi
   \else
     xxx 
   \fi}
 \gdef\xxx{\@ifnextchar[\xxx@lab\xxx@nolab}
 \long\gdef\xxx@lab[#1]#2{{\bf [\xxxmark #2 ---{\sc #1}]}}
 \long\gdef\xxx@nolab#1{{\bf [\xxxmark #1]}}
}

\def\R{\mathbb{R}}

\def\eps{\epsilon}

\begin{document}

\title{\Large Lower Bounds for Compressed Sensing with Generative Models }

\author{
 	Akshay Kamath\\
 	UT Austin\\\texttt{kamath@cs.utexas.edu} 
 	\and
 	Sushrut Karmalkar\\
 	UT Austin\\\texttt{sushrutk@cs.utexas.edu} 
 	\and
	Eric Price\\
 	UT Austin\\\texttt{ecprice@cs.utexas.edu}}
\date{}
\maketitle

\begin{abstract}



  The goal of compressed sensing is to learn a structured signal $x$
  from a limited number of noisy linear measurements $y \approx Ax$.  In
  traditional compressed sensing, ``structure'' is represented by
  sparsity in some known basis.  Inspired by the success of deep
  learning in modeling images, recent work starting with~\cite{BJPD17}
  has instead considered structure to come from a generative model
  $G: \R^k \to \R^n$.  We present two results establishing the
  difficulty of this latter task, showing that existing bounds are
  tight.

  First, we provide a lower bound matching the~\cite{BJPD17} upper
  bound for compressed sensing from $L$-Lipschitz generative models
  $G$.  In particular, there exists such a function that requires
  roughly $\Omega(k \log L)$ linear measurements for sparse recovery
  to be possible.  This holds even for the more relaxed goal of
  \emph{nonuniform} recovery.

  Second, we show that generative models generalize sparsity as a
  representation of structure.  In particular, we construct a
  ReLU-based neural network $G: \R^{2k} \to \R^n$ with $O(1)$ layers
  and $O(kn)$ activations per layer, such that the range of $G$
  contains all $k$-sparse vectors.

\end{abstract}

\section{Introduction}

In compressed sensing, one would like to learn a structured signal
$x \in \R^n$ from a limited number of linear measurements
$y \approx Ax$.  This is motivated by two observations: first, there
are many situations where linear measurements are easy, in settings as
varied as streaming algorithms, single-pixel cameras, genetic testing,
and MRIs.  Second, the unknown signals $x$ being observed are
structured or ``compressible'': although $x$ lies in $\R^n$, it would
take far fewer than $n$ words to describe $x$.  In such a situation,
one can hope to estimate $x$ well from a number of linear measurements
that is closer to the size of the \emph{compressed representation} of
$x$ than to its ambient dimension $n$.

In order to do compressed sensing, you need a formal notion of how
signals are expected to be structured.  The classic answer is to use
\emph{sparsity}.  Given linear measurements\footnote{The algorithms we
  discuss can also handle post-measurement noise, where $y = Ax + \eta$.  We
  remove this term for simplicity: this paper focuses on lower bounds,
  and handling this term could only make things harder.} $y = Ax$ of
an arbitrary vector $x \in \R^n$, one can hope to recover an estimate
$x^*$ of $x$ satisfying
\begin{align}\label{eq:goalsparse}
  \norm{}{x - x^*} \leq C \min_{\text{$k$-sparse $x'$}}\norm{}{x - x'}
\end{align}
for some constant $C$ and norm $\norm{}{\cdot}$.  In this paper, we
will focus on the $\ell_2$ norm and achieving the guarantee with $3/4$
probability.  Thus, if $x$ is well-approximated by a $k$-sparse vector
$x'$, it should be accurately recovered.  Classic results such
as~\cite{CRT06} show that~\eqref{eq:goalsparse} is achievable when $A$
consists of $m = O(k \log \frac{n}{k})$ independent Gaussian linear
measurements.  This bound is tight, and in fact no distribution of
matrices with fewer rows can achieve this guarantee in either $\ell_1$
or $\ell_2$~\cite{DIPW}.

Although compressed sensing has had success, sparsity is a limited
notion of structure.  Can we learn a richer model of signal structure
from data, and use this to perform recovery?  In recent years, deep
convolutional neural networks have had great success in producing rich
models for representing the manifold of images, notably with
generative adversarial networks (GANs)~\cite{GANs} and variational
autoencoders (VAEs)~\cite{VAEs}.  These methods produce generative
models $G: \R^k \to \R^n$ that allow approximate sampling from the
distribution of images.  So a natural question is whether these
generative models can be used for compressed sensing.

In~\cite{BJPD17} it was shown how to use generative models to achieve
a guarantee analogous to~\eqref{eq:goalsparse}: for any $L$-Lipschitz
$G: \R^k \to \R^n$, one can achieve
\begin{align}\label{eq:goalG}
  \norm{2}{x - x^*} \leq C \min_{z' \in B_k(r)}\norm{2}{x - G(z')} + \delta,
\end{align}
where $r, \delta > 0$ are parameters, $B_k(r)$ denotes the
radius-$r$ $\ell_2$ ball in $\R^k$ and Lipschitzness is defined with respect to 
the $\ell_2$-norms, using only $m = O(k \log \frac{Lr}{\delta})$ measurements.  
Thus, the recovered vector is almost as good as the nearest point in the
\emph{range of the generative model}, rather than in the set of
$k$-sparse vectors. We will refer to the problem of achieving the
guarantee in ~\eqref{eq:goalG} as ``function-sparse recovery''.

Our main theorem is that the~\cite{BJPD17} result is tight: for any
setting of parameters $n, k, L, r, \delta$, there exists an
$L$-Lipschitz function $G: \R^k \to \R^n$ such that any algorithm
achieving~\eqref{eq:goalG} with $3/4$ probability must have
$\Omega(\min(k \log \frac{Lr}{\delta}, n))$ linear measurements.
Notably, the additive error $\delta$ that was unnecessary in sparse
recovery is necessary for general Lipschitz generative model recovery.

A concurrent paper \cite{LiuScar} proves a lower bound for a restricted version 
of \eqref{eq:goalG}. They show a lower bound when the vector that $x$ lies in 
the image of $G$ and for a particular value of $\delta$. Our results, in 
comparison, apply to the most general version of the problem and are proven 
using a simpler communication complexity technique.

The second result in this paper is to directly relate the two notions
of structure: sparsity and generative models.  We produce a simple
Lipschitz neural network $G_{sp}: \R^{2k} \to \R^n$, with ReLU activations,
$2$ hidden layers, and maximum width $O(kn)$, so that the range of $G$
contains all $k$-sparse vectors.

A second result of~\cite{BJPD17} is that for ReLU-based neural
networks, one can avoid the additive $\delta$ term and achieve a
different result from~\eqref{eq:goalG}:
\begin{align}\label{eq:goalG2}
  \norm{2}{x - x^*} \leq C \min_{z' \in \R^k}\norm{2}{x - G(z')}
\end{align}
using $O(k d \log W)$ measurements, if $d$ is the depth and $W$ is the
maximum number of activations per layer.  Applying this result to our
sparsity-producing network $G_{sp}$ implies, with $O(k \log n)$
measurements, recovery achieving the standard sparsity
guarantee~\eqref{eq:goalsparse}.  So the generative-model
representation of structure really is more powerful than sparsity.

\section{Proof overview}

As described above, this paper contains two results: an
$\Omega(\min(k \log \frac{Lr}{\delta}, n))$ lower bound for compressed
sensing relative to a Lipschitz generative model, and an $O(1)$-layer
generative model whose range contains all sparse vectors.  These
results are orthogonal, and we outline each in turn. 

\subsection{Lower bound for Lipschitz generative recovery.}  
Over the last decade, lower bounds for sparse recovery have been
studied extensively.  The techniques in this paper are most closely
related to the techniques used in \cite{DIPW}.


Similar to \cite{DIPW}, our proof is based on communication complexity. We will 
exhibit an $L$-Lipschitz function $G$ and a large finite set $Z \subset Im(G) 
\subset B_n(R)$ of points that are well-separated. Then, given a point $x$ that 
is picked uniformly at random from $Z$, we show how to identify it
from $Ax$ using the function-sparse recovery algorithm. This implies $Ax$ also 
contains a lot of information, so $m$ must be fairly large.

Formally, we produce a generative model whose range includes a large, 
well-separated set:
\begin{theorem}\label{thm:mapping}
	Given $R > 0$  satisfying $R> 2Lr$, there exists an $O(L)-$Lipschitz 
	function $G:\R^k \rightarrow \R^n$, and $X\subseteq B_k(r)$ 	 
	such that 	
	\begin{enumerate}
		\item[(1)] for all $x\in X$, $G(x)\in \{\pm\frac{R}{\sqrt{n}}\}^n$
		\item[(2)] for all $x \in X$, $\norm{2}{G(x)}\leq R$
		\item[(3)] for all $x,y \in X$, 
		$\norm{2}{G(x)-G(y)}\geq\frac{R}{\sqrt{6}} $
		\item[(4)] 
		$\log(\abs{X})=\Omega\left(\min(k\log(\frac{Lr}{R})), 
		n\right)$
	\end{enumerate}                    
\end{theorem}  

Now, suppose we have an algorithm that can perform function-sparse recovery 
with respect to $G$ from Theorem \ref{thm:mapping}, with approximation factor 
$C$, and error $\delta<R/8$ within the radius $r$ ball in $k$-dimensions. Set
$t = \Theta(\log n)$, and for any $z_1, z_2, \dotsc, z_t \in Z = G(X)$ take
\[
z = \eps^t z_1 + \eps^{t-1} z_2 + \eps^{t-1} z_3 + \dotsc + z_t
\]
for $\eps = \frac{1}{8(C+1)}$ a small constant.  The idea of the
proof is the following: given $y = Az$, we can recover $\widehat{z}$ such
that
\[
\norm{2}{\widehat{z} - z_1} \leq \norm{2}{z - z_1} + \norm{2}{\widehat{z} - z} 
+ \delta \leq (C+1)\norm{2}{z - z_1} + \delta  < 
R/8 + R/8 = R/4
\]
and so, because $Z$ has minimum distance $R/\sqrt{6}$, we can exactly
recover $z_t$ by rounding $\widehat{z}$ to the nearest element of $Z$.
But then we can repeat the process on $(Az - Az_t)$ to
find $z_{t-1}$, then $z_{t-2}$, up to $z_1$, and learn 
$t \lg \abs{Z} = \Omega(t k \log(Lr/R))$ bits total.  Thus
$Az$ must contain this many bits of information; but if the entries of $A$ are
rational numbers with $\text{poly}(n)$ bounded numerators and
(the same) $\text{poly}(n)$ bounded denominator, then each entry of $Az$ can be described in
$O(t + \log n)$ bits, so
\[
m \cdot O(t + \log n) \geq \Omega(t k \log(Lr/R))
\]
or $m \geq \Omega(k \log(Lr/R))$.

There are two issues that make the above outline not totally
satisfactory, which we only briefly address how to resolve here.
First, the theorem statement makes no supposition on the entries of
$A$ being polynomially bounded.  To resolve this, we perturb $z$
with a tiny (polynomially small) amount of additive Gaussian noise,
after which discretizing $Az$ at an even tinier (but still
polynomial) precision has negligible effect on the failure
probability.  The second issue is that the above outline requires
the algorithm to recover all $t$ vectors, so it only applies if the
algorithm succeeds with $1 - 1/t$ probability rather than
constant probability.  This is resolved by using a reduction from
the \emph{augmented indexing} problem, which is a one-way
communication problem where Alice has $z_1, z_2, \dotsc, z_t \in Z$,
Bob has $i \in [Z]$ and $z_{i+1}, \cdots, z_n$, and Alice must send Bob a 
message so that Bob can output $z_i$ with $2/3$ probability.  This still 
requires $\Omega(t \log \abs{Z})$ bits of communication, and can be solved in
$O(m (t +\log n))$ bits of communication by sending $Az$ as above.  
Formally, our lower bound states:
\begin{theorem}\label{thm:discrete-randomized}
	Consider any $L, r, \delta$ with $\delta \leq Lr/4$.  There exists
	an $L$-Lipschitz function $G^*:\R^k \rightarrow \R^n$ such that, if
	$\mathcal{A}$ is an algorithm which picks a matrix
	$A\in \R^{m\times n}$, and given $Ax$ returns an $x^*$
	satisfying~\eqref{eq:goalG} with probability $\geq 3/4$, then
	$m = \Omega(\min(k\log(Lr/\delta), n))$.
\end{theorem}

\paragraph{Constructing the set.}  The above lower bound approach, relies on 
finding a large, well-separated set $Z$ as in Theorem \ref{thm:mapping}.

We construct this aforementioned set $Z$ within the $n$-dimensional $\ell_2$ 
ball of radius $R$ such that any two points in the set are at least $\Omega(R)$ 
apart. Furthermore, since we wish to use a function-sparse recovery 
algorithm, we describe a function $G:\R^{k}\rightarrow \R^{n}$ and set the 
radius $R$ such that $G$ is $L$-Lipschitz. In order to get the desired lower 
bound, the image of $G$ needs to contain a subset of at least 
$(Lr)^{\Omega(k)}$ points.

First, we construct a mapping as described above from $\R$ to $\R^{n/k}$ i.e we 
need to find $(Lr)^{\Omega(k)}$ points in $B_{n/k}(R)$ that are 
mutually far apart. We show that certain binary linear codes over the alphabet 
$\{ \pm R/\sqrt{n} \}$ yield such points that are mutually $R/\sqrt{3k}$ apart. 
We construct a $O(L)$-Lipschitz mapping of $O(\sqrt{Lr})$ points in the 
interval $[0, r/\sqrt{k}]$ to a subset of these points.  

In order to extend this construction to a mapping from $\R^k$ to $\R^n$, we 
apply the above function in a coordinate-wise manner. This would result in a 
mapping with the same Lipschitz parameter. The points in $\R^n$ that are images 
of these points lie in a ball of radius $R$ but could potentially be 
$R/\sqrt{3k}$ close. To get around this, we use an error correcting code over a 
large alphabet to choose a subset of these points that is large enough and such 
that they are still mutually $R/\sqrt{6}$ far apart. 

\subsection{Sparsity-producing generative model.}
To produce a generative model whose range consists of all $k$-sparse
vectors, we start by mapping $\R^2$ to the set of positive $1$-sparse
vectors.  For any pair of angles $\theta_1, \theta_2$, we can use a
constant number of unbiased ReLUs to produce a neuron that is only active at 
points whose representation $(r, \theta)$ in polar coordinates has $\theta \in 
(\theta_1, \theta_2)$.  Moreover, because unbiased ReLUs behave linearly, the 
activation can be made an
arbitrary positive real by scaling $r$ appropriately.  By applying
this $n$ times in parallel, we can produce $n$ neurons with disjoint
activation ranges, making a network $\R^2 \to \R^n$ whose range
contains all $1$-sparse vectors with nonnegative coordinates.

By doing this $k$ times and adding up the results, we produce a
network $\R^{2k} \to \R^n$ whose range contains all $k$-sparse vectors
with nonnegative coordinates.  To support negative coordinates, we
just extend the $k=1$ solution to have two ranges within which it is non-zero: 
for one range of $\theta$ the output is positive, and for another the output is 
negative.

This results in the following theorem:
\begin{theorem}\label{thm:k-sparse}
	There exists a 2 layer neural network $G_{sp}:\R^{2k} \rightarrow \R^n$ with 
	width $O(nk)$ such that $\{x \mid \norm{0}{x} = k \} \subseteq Im(G)$
\end{theorem}

\section{Lower bound proof}

In this section, we prove a lower bound for the sample complexity of 
function-sparse recovery by a reduction from a communication game. We show 
that  the communication game can be won by sending a vector $Ax$ and then 
performing function-sparse recovery. A lower bound on the communication 
complexity of the game implies a lower bound on the number of bits used to 
represent $Ax$ if $Ax$ 
is discretized. We can then use this to lower bound the number of measurements 
in $A$. 

Since we are dealing in bits in the communication game and 
the entries of a sparse recovery matrix can be arbitrary reals, we will need to discretize each measurement. We show first that discretizing the measurement matrix by rounding does not change the resulting measurement too much and will 
allow for our reduction to proceed.  

\paragraph{Notation.}
We use $B_k(r) = \{x\in \R^{k} \mid \norm{2}{x}\leq r \}$ to denote the 
$k$-dimensional ball of radius $r$. Given a function $g:\R^a \rightarrow \R^b$, 
$g^{\otimes k}:\R^{ak} \rightarrow \R^{bk}$ denotes a function that  the maps a 
point $(x_1, \dotsc, x_{ak})$ to $(g(x_1, \dotsc, x_a), g(x_{a+1},\dotsc, 
x_{2a}), 
\dotsc, g(x_{a(k-1)+1}, \dotsc, x_{ak}))$. For any function $G : A \rightarrow 
B $, we use $Im(G)$ to denote $\{ G(x) \mid x \in A \}$.

\paragraph{Matrix conditioning.}  We first show that, without loss of
generality, we may assume that the measurement matrix $A$ is
well-conditioned.  In particular, we may assume that the rows of $A$
are orthonormal.

We can multiply $A$ on the left by any invertible matrix to get
another measurement matrix with the same recovery characteristics.  If
we consider the singular value decomposition $A = U\Sigma V^*$, where
$U$ and $V$ are orthonormal and $\Sigma$ is 0 off the diagonal, this
means that we can eliminate $U$ and make the entries of $\Sigma$ be
either $0$ or $1$.  The result is a matrix consisting of $m$
orthonormal rows.

\paragraph{Discretization.}  For well-conditioned matrices $A$, we use
the following lemma (similar to one from \cite{DIPW}) to show that we
can discretize the entries without changing the behavior by much:

\begin{lemma}\label{thm:discretizing}
  Let $A\in \R^{m\times n}$ be a matrix with orthonormal rows.  Let $A'$
  be the result of rounding $A$ to $b$ bits per entry.  Then for any
  $v \in \mathbb{R}^n$ there exists an $s \in \mathbb{R}^n$ with $A'v
  = A(v - s)$ and $\norm{2}{s} < n 2^{-b}\norm{2}{v}$.
\end{lemma}
\begin{proof}
  Let $A'' = A - A'$ be the error when discretizing $A$ to
  $b$ bits, so each entry of $A''$ is less than $2^{-b}$. Then for
  any $v$ and $s = A^TA''v$, we have $As = A''v$ and
  \begin{align*}
    \norm{2}{s} &= \norm{2}{A^TA''v} \leq \norm{2}{A''v}\\ &\leq
    m 2^{-b}\norm{2}{v} \leq n 2^{-b}\norm{2}{v}.
  \end{align*}
\end{proof}

\paragraph{The Augmented Indexing problem.}
As in \cite{DIPW}, we use the {\sf Augmented Indexing} communication game 
which is defined as follows: There are two parties, Alice and Bob. Alice is 
given a string $y \in \{0,1\}^d$. Bob is given an index $i \in [d]$, together 
with $y_{i+1}, y_{i+2}, \ldots, y_d$. The parties also share an arbitrarily 
long common random string $r$. Alice sends a single message $M(y,r)$ to Bob, 
who must output $y_i$ with probability at least $2/3$, where the probability is 
taken over $r$. We refer to this problem as {\sf Augmented Indexing}. The 
communication cost of {\sf Augmented Indexing} is the minimum, over all correct 
protocols, of length $\abs{M(y,r)}$ on the worst-case choice of 
$r$ and $y$.

The following theorem is well-known and follows from Lemma 13 of \cite{MNSW98} 
(see, for example, an explicit proof in \cite{DIPW})
\begin{theorem}\label{thm:AIND}
The communication cost of {\sf Augmented Indexing} is $\Omega(d)$.
\end{theorem}

\paragraph{A well-separated set of points.}
We would like to prove Theorem~\ref{thm:mapping}, getting a large set
of well-separated points in the image of a Lipschitz generative model.
Before we do this, though, we prove a $k=1$ analog:
\begin{lemma}\label{lem:path}
	There is a set of points $P$ in $B_n(1) \subset \mathbb{R}^n$ of size 
	$2^{\Omega(n)}$ such that for each pair of points $x, y \in P$
	\[ \|x - y\| \in \left[ \sqrt{ \frac{1}{3}}, \sqrt{\frac{2}{3}} \right] 
	\]
\end{lemma} 

\begin{proof}
	Consider a $\tau$-balanced linear code over the alphabet $\{ \pm 
	\frac{1}{\sqrt{n}} \}$ with message length $M$. It is known that such 
	codes exist with block length $O(M/\tau^2)$ \cite{TB09}. Setting the 
	block length to be $n$ and $\tau = 1/6$, we get that there is a set of 
	$2^{\Omega(n)}$ points in $\R^n$ such that the pairwise hamming 
	distance is between $\left[ \frac{n}{3}, \frac{2n}{3}\right]$, i.e. the 
	pairwise $\ell_2$ distance is between $\left[ \sqrt{ \frac{1}{3}}, 
	\sqrt{\frac{2}{3}} \right]$. 
\end{proof}

Now we wish to extend this result to arbitrary $k$ while achieving the 
parameters in Theorem \ref{thm:mapping}.

\begin{proof}[Proof of Theorem \ref{thm:mapping}]
  We first define an $O(L)$-Lipschitz map $g: \mathbb{R} \rightarrow 
  \mathbb{R}^{n/k}$ that goes through a set of points that are pairwise 
  $\Theta\left( \frac{R}{\sqrt{k}}\right)$ apart. Consider the set of points 
  $P$ from Lemma~\ref{lem:path} scaled to $B_{n/k}(\frac{R}{\sqrt{k}})$. 
%
%
  Observe that $\abs{P}\geq  \exp \left(\Omega \left(n/k\right) \right) \geq 
  \min \left(\exp\left(\Omega \left(n/k\right) \right), Lr/R \right)$. 
  Choose subset  $P'$ that such that it contains exactly 
  $\min \left(Lr/R, \exp(\Omega(n/k)) \right)$ points and let $g_1:[0, 
  r/\sqrt{k}] 
  \rightarrow P'$ be a piecewise linear function that goes through all the 
  points in $P'$ in order. Then, we define $g: 
  \R\rightarrow \R^{n/k}$ as:
  \[
     g(x) = \begin{cases}
	     g_1(0) &\quad \text{if } x<0\\
	     g_1(x) &\quad \text{if } 0\leq x\leq r/\sqrt{k}\\
	     g_1(\frac{R}{\sqrt{k}})  &\quad \text{if } x\geq r/\sqrt{k}
	     \end{cases}
  \] 
  Let $I = \{\frac{r}{\sqrt{k} \abs{P'}}, \dots, 
 \frac{r}{\sqrt{k}} \}$ be the points that are pre-images of elements of 
  $P'$. Observe that $g$ is $O(L)$-Lipschitz since within the interval $[0, 
  r/\sqrt{k}]$, since it maps each interval of length $\frac{r}{\sqrt{k} 
  \abs{P'}} \geq \frac{rR}{\sqrt{k} L r } = \frac{R}{L\sqrt{k}}$ to an 
  interval of length at most $O(R/\sqrt{k})$. 
	
  Now, consider the function  $G := g^{\otimes k}:\R^k \rightarrow  \R^{n}$. 
  Observe that $G$ is also $O(L)$ Lipschitz, 
  \begin{align*}
	\norm{2}{G(x_1, \dotsc, x_k) - G(y_1, \dotsc, y_k)}^2 & = \sum_{i\in
	  [k]} \norm{2}{g(x_i) - g(y_i)}^2 \\
	& \leq \sum_{i\in [k]} O(L^2)\norm{2}{x_i - y_i}^2\\
	& = O(L^2)\norm{2}{x-y}^2
  \end{align*}
  Also, for every point $(x_1, \dotsc, x_k) \in I^{k}$, $\norm{2}{G(x_1, 
  \dotsc, x_k)} = \sqrt{\sum_{i\in[k]} \norm{2}{g(x_i)}^2} \leq R$. However, 
  there still exist distinct points $x,y\in I^{k}$(for instance points that 
  differ at exactly one coordinate) such that $\norm{2}{G(x)-G(y)}\leq 
  O(\frac{R}{\sqrt{k}})$.
  
  We construct a large subset of the points in $I^{k}$ such that any two 
  points   in this subset are far apart using error correcting codes. Consider 
  the $A \subset P'$ s.t. $\abs{A}>\abs{P'}/2$ is a prime. For any integer $z > 
  0$, there is a prime between $z$ and $2z$, so such a set $A$ exists. Consider 
  a Reed-Solomon code of block length $k$, message length $k/2$, distance $k/2$ 
  and alphabet $A$. The existence of such a code implies that there is a subset 
  $X'$ of $(P')^k$ of size at least  
  $(\abs{P'}/2)^{k/2}$ such that every pair of distinct elements from this set 
  disagree in $k/2$  coordinates. 
  
  This translates into a distance of $\frac{R}{\sqrt{6}}$ in 
  2-norm. So, if we set $G = g^{\otimes k}$ and $X\subset I^k$ to $G^{-1}(X')$, 
  we get have a set of $(\abs{P'}/2)^{k/2}\geq (\min(\exp(\Omega(n/k)), 
  Lr/R))^{k/2}$ points which are $\frac{R}{\sqrt{6}}$ apart in  
  2-norm, lie within  the $\ell_2$ ball of radius $R$.
\end{proof}


\paragraph{Lower bound.}  We now prove the lower bound for function-sparse recovery.

\begin{proof}[Proof of Theorem \ref{thm:discrete-randomized}.]
	
	An application of Theorem \ref{thm:mapping} with $R = \sqrt{Lr\delta}$ 
	gives us a set of points $Z$ and $G$ such that $Z = G(X)\subseteq 
	\R^{n}$ such that  $\log(\abs{Z}) = 
	\Omega(\min(k\log(\frac{Lr}{\delta}), n))$, and 
	for all $x\in Z$,  $\norm{2}{x}\leq \sqrt{Lr\delta}$ and for all $x,x' \in Z$, 
	$\norm{2}{x-x'}\geq \sqrt{Lr\delta}/\sqrt{6}$. 
	Let $d = \floor{\log\abs{X}} \log n$, and let $D = 16\sqrt{3}(C+1)$.

We will show how to solve the {\sf Augmented Indexing} problem on instances of size $d = \log(\abs{Z})\cdot \log(n) = \Omega(k \log (Lr) \log n)$ with communication 
cost $O(m \log n)$. The theorem will then follow by Theorem \ref{thm:AIND}.

Alice is given a string $y \in \{0,1\}^d$, and Bob is given $i \in
[d]$ together with $y_{i+1}, y_{i+2}, \ldots, y_d$, as in the setup
for {\sf Augmented Indexing}.

Alice splits her string $y$ into $\log n$ contiguous chunks $y^1, y^2, \ldots, y^{\log n}$, each containing $\floor{\log \abs{X}}$ bits. She uses $y^j$ as an index into the set $X$ to choose $x_j$.  Alice defines
\[
x = D^1 x_1 + D^2 x_2 + \cdots + D^{\log n} x_{\log n}.
\]
Alice and Bob use the common randomness $\mathcal{R}$ to agree upon a random matrix $A$ with orthonormal rows.  Both Alice and Bob round $A$ to form $A'$ with $b = \Theta(\log(n))$ bits per entry. Alice computes $A'x$ and transmits it to Bob. Note that, since $x \in \left\{ \pm \frac{1}{\sqrt{n}} \right\}$ the $x$'s need not be discretized. 

From Bob's input $i$, he can compute the value $j=j(i)$ for which the bit $y_i$ occurs in $y^{j}$. Bob's input also contains $y_{i+1},
\ldots, y_n$, from which he can reconstruct $x_{j+1}, \ldots, x_{\log n}$, and in particular can compute
\[
z = D^{j+1}x_{j+1} + D^{j+2}x_{j+2} + \cdots + D^{\log n}x_{\log n}.
\]
Set $w = \frac{1}{D^j} (x-z) = \frac{1}{D^j} \sum_{i=1}^j 
D^{i}x_i$.  Bob then computes $A'z$, and
using $A'x$ and linearity, he can compute $\frac{1}{D^j}\cdot A'(x-z) = 
A'w$.  Then
\[
\norm{2}{w} \leq \frac{1}{D^j} \sum_{i=1}^j R\cdot D^{i} < R.
\]
So from Lemma~\ref{thm:discretizing}, there exists some $s$ with $A'w
= A(w-s)$ and
\[
\norm{2}{s} < n^2 2^{-b}\norm{2}{w} < \frac{R}{D^j n^2}.
\]
Ideally, Bob would perform recovery on the vector $A(w-s)$ and show that the 
correct point $x_j$ is recovered. However, since $s$ is correlated with $A$ and 
$w$, Bob needs to use a slightly more complicated technique. 

Bob first chooses another vector $u$ uniformly from $B_n(R/D^{j})$ and computes $A(w 
- s - u) = A'w - Au$. He then runs the estimation algorithm $\mathcal{A}$ on $A$ and $A(w-s-u)$,
obtaining $\hat{w}$. We have that $u$ is independent of $w$ and $s$, and that 
$\norm{2}{u} \leq\frac{R}{D^j}(1 - 1/n^2) \leq \frac{R}{D^j} - \norm{2}{s}$ 
with probability $\frac{\text{Vol}(B_n(\frac{R}{D^j}(1 - 
1/n^2)))}{\text{Vol}(B_n(\frac{R}{D^j}))} = (1 - 1/n^2)^n > 1 - 1/n$.  But 
$\{w - u \mid \norm{2}{u} \leq \frac{R}{D^j} - \norm{2}{s}\} \subseteq \{w - s 
- u \mid \norm{2}{u} \leq \frac{R}{D^j}\}$, so as a distribution over $u$, the 
ranges of the random variables $w - s - u$ and $w - u$ overlap in at least a $1 
- 1/n$ fraction of their volumes. Therefore $w-s-u$ and $w-u$ have statistical 
distance at most $1/n$.  The distribution of $w - u$ is independent of $A$, so 
running the recovery algorithm on $A(w - u)$ would work with probability at
least $3/4$. Hence with probability at least $3/4 - 1/n\geq 2/3$ (for $n$ large 
enough), $\hat{w}$ satisfies the recovery criterion for $w - u$, meaning
\[
\norm{2}{w-u-\hat{w}} \leq C \min_{w'\in Im(G)} \norm{2}{w-u-w'} + \delta
\]
Now,
\begin{align*}
  \norm{2}{x_j - \hat{w}} &\leq  \norm{2}{w-u - x_j} +  \norm{2}{w-u - 
  \hat{w}}\\
  &\leq (1 + C) \norm{2}{w-u -x_j} + \delta\\
  &\leq (1 + C) \left(\norm{2}{u} +   \frac{1}{D^j}\cdot \sum_{i=1}^{j-1} 
  \norm{2}{D^ix_i} \right) + \delta \\
  &\leq 2(1 + C)R/D + \delta\\
  &< R \cdot \frac{2(1+C)}{D} +\delta \\
  &= \frac{1}{8\sqrt{3}} \cdot R + \delta.
\end{align*}
Since $\delta < Lr/4$, this distance is strictly bounded by $R/2\sqrt{6}$. 
Since the minimum distance in $X$ is $R/\sqrt{6}$, this means $\norm{2}{D^jx_j 
- \hat{w}} < \norm{2}{D^jx' - \hat{w}}$ for all $x'\in X, x' \neq x_j$. So Bob 
can correctly identify $x_j$ with probability at least $2/3$.  From $x_j$ he 
can recover $y^j$, and hence the bit $y_i$ that occurs in $y^j$.

Hence, Bob solves {\sf Augmented Indexing} with probability at least
$2/3$ given the message $A'x$.  Each entry of $A'x$ takes $O(\log n)$ bits to 
describe because $A'$ is discretized to up to $\log(n)$ bits and $x\in \{ 
\pm\frac{1}{\sqrt{n}} \}^n $. Hence, the communication cost of this protocol is 
$O(m \cdot \log n)$. By Theorem \ref{thm:AIND}, $m \log n = \Omega(\min(k 
\log(Lr/\delta) , n) \cdot \log n)$, or $m = \Omega(\min(k \log (Lr/\delta), 
n))$.
\end{proof}
\section{Reduction from $k$-sparse recovery}
We show that the set of all $k$-sparse vectors in $\R^n$ is contained in the 
image of a 2 layer neural network. This shows that function-sparse recovery is 
a generalization of sparse recovery.

\begin{lemma}\label{lem:1-sparse}
	There exists a 2 layer neural network $G:\R^2 \rightarrow \R^n$ with 
	width $O(n)$ such that $\{ x \mid \norm{0}{x}=1 \} \subseteq Im(G)$ 
\end{lemma}

Our construction is intuitively very simple. We define two gadgets $G^+_i$ and 
$G^-_i$. $G^+_i\geq 0$ and $G^+_i(x_1,x_2)\neq0$ iff $\arctan(x_2/x_1) \in 
[i\cdot \frac{2\pi}{n}, (i+1) \cdot \frac{2\pi}{n} ]$. Similarly 
$G^-_i(x_1,x_2)\leq 0$ and $G^-_i(x_1,x_2)\neq 0$ iff 
$\arctan(x_2/x_1) \in [\pi + i\cdot \frac{2\pi}{n}, \pi + (i+1) \cdot 
\frac{2\pi}{n}]$. Then, we set the $i^{\text{th}}$ output node $(G(x_1, 
x_2))_i = G^+_i(x_1, x_2) + G^-_i(x_1, x_2)$. Varying the distance of $(x_1, 
x_2)$ from the origin will allow us to get the desired value at the output node 
$i$.

\begin{proof}
Let $\alpha= \frac{\pi}{n+1}$. Let $[x]_+ = x\cdot \mathbb{I}(x\geq 0)$ denote 
the unbiased ReLU function that preserves positive values and  $[x]_- = x\cdot 
\mathbb{I}(x\leq 0)$ denote the unbiased ReLU function that 
preserves negative values. We define $G_i^+:\R^2\rightarrow \R$ as follows:

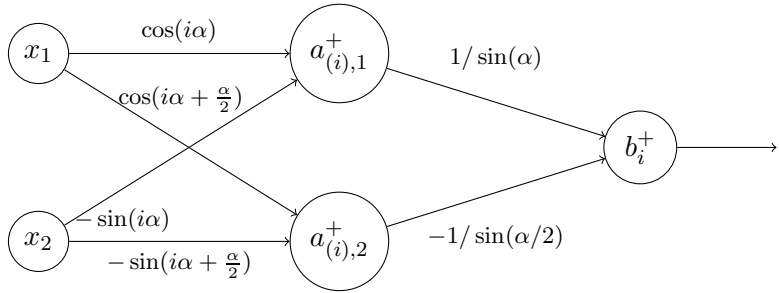
\begin{center}

\begin{tikzpicture}
\tikzstyle{place}=[circle, draw=black, minimum size = 8mm]
\draw node at (0, -1.25) [place] (first_1) {$x_1$};
\draw node at (0, -3*1.25) [place] (first_2) {$x_2$};	

\draw node at (4, -1.25) [place] (second_1){$a^+_{(i),1}$};
\draw node at (4, -3*1.25) [place] (second_2){$a^+_{(i),2}$};

\draw node at (8, -2.5) [place] (fourth_2){$b^+_i$};

\draw node at (10,  -2.5) [circle, ] (output_2){};

\draw [->] (first_1) -- node[midway, above] {\footnotesize $\cos(i\alpha)$} 
++ (second_1) ;
\draw [->] (first_1)  -- node[midway, above = 0.27cm] {\footnotesize  
$\cos(i\alpha 
+ \frac{\alpha}{2})$} ++ (second_2);
\draw [->] (first_2) -- node[near start, below = 0.12cm] {\footnotesize 
$-\sin(i\alpha)$} ++ (second_1);
\draw [->] (first_2) -- node[midway, below] {\footnotesize $-\sin(i\alpha + 
\frac{\alpha}{2})$} ++ (second_2);

\draw [->] (second_1) -- node[midway, above=0.3cm] {\footnotesize 
$1/\sin(\alpha)$} 
++ (fourth_2);
\draw [->] (second_2) -- node[midway, below=0.3cm] {\footnotesize 
$-1/\sin(\alpha/2)$} 
++ (fourth_2);

\draw [->] (fourth_2) to (output_2);

\end{tikzpicture}
\end{center}
$G^+_i$ is a 2 layer neural network gadget that produces positive values at 
output node $i$ of $G$. We define each of the hidden nodes of the neural 
network $G^+_i$ as follows:
\begin{align*}
a^+_{(i),1} &= \Big[ \cos(i\alpha)x_1 - \sin(i\alpha) x_2  \Big]_{+}\\
a^+_{(i),2} &= \Big[ \cos\big(i\alpha + \frac{\alpha}{2}\big)x_1 - 
\sin\big(i\alpha + 
\frac{\alpha}{2}\big) x_2  \Big]_{+}\\
b^+_{(i)} &= \Big[\frac{a^+_{(i),1}}{\sin(\alpha)}  - 
\frac{a^+_{(i),2}}{\sin(\alpha/2)}  \Big]_{+}
\end{align*}

In a similar manner, $G^-_i$ which produces negative values at 
output node $i$ of $G$ with the internal nodes defined as:
\begin{align*}
a^-_{(i),1} &= \Big[ \cos(\pi + i\alpha)x_1 - \sin(\pi + i\alpha) x_2  
\Big]_{+}\\
a^-_{(i),2} &= \Big[ \cos\big(\pi + i\alpha + \frac{\alpha}{2}\big)x_1 - 
\sin\big(\pi + i\alpha + 
\frac{\alpha}{2}\big) x_2  \Big]_{+}\\
b^-_{(i)} &= \Big[ 
\frac{a^-_{(i),2}}{\sin(\alpha/2)} - \frac{a^-_{(i),1}}{\sin(\alpha)}\Big]_{-}
\end{align*}
The last ReLU activation preserves only negative values. Since $G^+_i$ and 
$G^-_i$ are identical up to signs in the second hidden layer, we only analyze 
$G^+_i$'s. \\
Consider $i\in [n]$. Let $\beta = i\alpha$ and $(x_1, x_2) = (t\sin(\theta), 
t\cos(\theta))$. Then using the identity $\sin(A)\cos(B) - \cos(A)\sin(B) = 
\sin(A-B)$,  
\begin{align*}
\cos(\beta)x_1 - \sin(\beta) x_2 & = t \big( \cos(\beta)\sin(\theta) - 
\sin(\beta) \cos(\theta) \big)\\
	& = t \sin(\theta -\beta)
\end{align*}
This is positive only when $\theta \in (\beta, \pi + \beta)$. 
Similarly, $\cos(\beta + \alpha/2)x_1 - \sin(\beta + \alpha/2) x_2 = t 
\sin(\theta - (\beta + \alpha/2))$ and is positive only when  $\theta \in 
(\beta + \alpha/2, \pi + \beta + \alpha/2)$. So, $a^+_{(i),1}$ and 
$a^+_{(i),2}$ are both non-zero when $\theta\in(\beta + \alpha/2, \pi + 
\beta)$. Using some elementary trigonometry, we may see that:
\begin{align*}
	\frac{a_1^{(i)}}{\sin(\alpha)} - \frac{a_2^{(i)}}{\sin(\alpha/2)} 
	 & = t\Big(\frac{\sin(\theta - \beta)}{\sin(\alpha)}  - 
	\frac{\sin(\theta - (\beta + 
		\frac{\alpha}{2}))}{\sin(\alpha/2)}\Big)\\
	& = \frac{t\sin(\beta -\theta + \alpha)}{\sin(\alpha/2)}   
\end{align*}
In Fact \ref{app:trig-id}, we show a proof of the above identity. 
Observe that when $\theta>\beta+\alpha$, this term is negative and 
hence $b^{i} = 0$. So, we may conclude that  $G^+_i((x_1, x_2))\neq 0$ if and 
only if $(x_1, x_2) = (t\sin(\theta), t\cos(\theta))$ with $\theta\in 
((i-1)\alpha, i\alpha)$. Also, observe that 
$G^+_i(t\sin(\beta + \alpha/2), t\cos(\beta + \alpha/2)) = t$. 
Similarly $G^-_i$ is non-zero only if and only if $\theta\in 
[\pi + i\alpha, \pi + (i+1)\alpha]$ and $G^-_i(t\sin(\pi + i\alpha + \alpha/2), 
t\cos(\pi + i\alpha + \alpha/2)) = -t$. Since $\alpha = \frac{\pi}{n+1}$, the 
intervals within which each of $G^+_1, \dotsc, G^+_n$ ,$G^-_1, \dotsc, 
G^-_n$ are non-zero do not intersect.

So, given a vector $z'$ such that $\norm{0}{z}=1$ with $z_{i'} \neq 0$, if 
$z_{i'} > 0$, set
\begin{align*}
 x_1 &= \abs{z_{i'}}\sin(i'\alpha + \alpha/2)\\ 
 x_2 &= \abs{z_{i'}}\cos(i'\alpha + \alpha/2)
\end{align*}
and if $z_{i'} < 0$, set 
\begin{align*}
x_1 &= \abs{z_{i'}}\sin(\pi + i'\alpha + \alpha/2)\\
x_2 &= \abs{z_{i'}}\cos(\pi + i'\alpha + \alpha/2)
\end{align*}
 Observe that: 
\[
G^+_{i'}((x_1, x_2)) + G^-_{i'}((x_1, x_2)) = z_{i'}  
\]
and for all $j\neq i'$
\[
G^+_{j}((x_1, x_2)) + G^-_{j}((x_1, x_2)) = 0  
\]
So, if $G(x) = (G^+_1(x) + G^-_1(x), \dotsc, G^+_n(x) + G^-_n(x))$, $G$ is a 
2-layer neural network with width $O(n)$ such that $\{ x\mid \norm{0}{x}=1\} 
\subseteq Im(G)$.
\end{proof}

\begin{proof}[Proof of Theorem \ref{thm:k-sparse}.]
	Given a vector $z$ that is non-zero at $k$ coordinates, let $i_1< i_2 < 
	\cdots	<i_k$ be the indices at which $z$ is non-zero. We may use copies 
	of $G$	from Lemma \ref{lem:1-sparse} to generate $1$-sparse vectors 
	$v_1, \dotsc, v_k$ such that $(v_j)_{i_j} = z_{i_j}$. Then, we add these 
	vectors to obtain $z$. It is clear that we only used $k$ copies of $G$ 
	to create $G_{sp}$. So, $G_{sp}$ can be represented by a neural network with 2 
	layers. 	
\end{proof}

Theorem \ref{eq:goalsparse} provides a reduction which uses only 2 layers. 
Then, using the algorithm from Theorem \ref{eq:goalG2}, we can recover the 
correct 
$k$-sparse vector using $O(kd\log (nk))$ measurements. Since $d = 4$ and $\leq 
n$, this requires only $O(k \log n)$ linear measurements to perform 
$\ell_2/\ell_2$ $(k,C)$-sparse recovery.
\bibliographystyle{alpha}
\bibliography{ref}
\appendix
\section{Trigonometric identity}
\begin{fact}\label{app:trig-id}
	\[
	\frac{\sin(\beta + \frac{\alpha}{2} - \theta)}{\sin(\alpha/2)} - 
	\frac{\sin(\beta - \theta)}{\sin(\alpha)} = 
	\frac{\sin(\beta -\theta + \alpha)}{\sin(\alpha/2)} 
	\]
\end{fact}
\begin{proof}
\begin{align*}
\frac{\sin(\beta + \frac{\alpha}{2} - \theta)}{\sin(\alpha/2)} - 
\frac{\sin(\beta - \theta)}{\sin(\alpha)}
&=\frac{\sin(\beta + \frac{\alpha}{2} - 
	\theta)\sin(\alpha) - \sin(\beta - 
	\theta)\sin(\alpha/2)}{\sin(\alpha)\sin(\alpha/2)}\\
&= \frac{\frac{1}{2}\big( \cos(\beta -\theta - \frac{\alpha}{2}) - \cos(\beta - 
\theta +\frac{3\alpha}{2}) - \cos(\beta -\theta - 
\frac{\alpha}{2}) + \cos(\beta - \theta +\frac{\alpha}{2}) \big) 
}{\sin(\alpha)\sin(\alpha/2)}\\
&= \frac{\cos(\beta - \theta 
	+\frac{\alpha}{2}) -  \cos(\beta - \theta +\frac{3\alpha}{2})}{2 
	\sin(\alpha)\sin(\alpha/2)}\\
& = \frac{ \sin(\beta -\theta + 
	\alpha)\sin(\alpha)}{ \sin(\alpha)\sin(\alpha/2)}\\
& = \frac{\sin(\beta -\theta + \alpha)}{\sin(\alpha/2)}
\end{align*}
where we use the identity that $\sin(A)\sin(B) = \frac{1}{2}[\cos(A-B) - 
\cos(A+B)]$
\end{proof}

\end{document}